\documentclass[french,english]{revtex4-1}
\usepackage[T1]{fontenc}
\usepackage[latin9]{inputenc}
\usepackage{array}
\usepackage{textcomp}
\usepackage{multirow}
\usepackage{amstext}
\usepackage{amssymb}

\makeatletter

\providecommand{\tabularnewline}{\\}

\usepackage[colorlinks = true,
            linkcolor = red,
            urlcolor  = blue,
            citecolor = blue,
            anchorcolor = blue]{hyperref}

\makeatother

\usepackage{babel}
\makeatletter
\addto\extrasfrench{%
   \providecommand{\fg}{\ifdim\lastskip>\z@\unskip\fi~\frqq}%
}

\makeatother
\begin{document}

\title{\texttt{DYNAMICAL NONCOMMUTATIVE GRAPHENE}}

\author{Ilyas Haouam}

\email{ilyashaouam@live.fr}

\selectlanguage{english}%

\address{Laboratoire de Physique Mathématique et de Physique Subatomique (LPMPS),
Université Frères Mentouri, Constantine 25000, Algeria}

\author{S Ali Alavi}

\email{s.alavi@hsu.ac.ir}

\selectlanguage{english}%

\address{Departmen of Physics, Hakim Sabzevari University, P.O. Box 397, Sabzevar,
Iran}
\begin{abstract}
{\normalsize{}We study graphene in a two-dimensional dynamical noncommutative
space in the presence of a constant magnetic field. The model is solved
using perturbation theory and to the second order of perturbation.
The energy levels of the system are calculated and the corresponding
eigenstates are obtained. For all cases, the energy shift depends
on the dynamical noncommutative parameter $\tau$. Using the accuracy
of energy measurement we put an upper bound on the noncommutativity
parameter $\tau$. In addition, we investigate some of the thermodynamic
quantities of the system at zero temperature limit and extreme relativistic
case, which reveals interesting differences between commutative and
dynamical noncommutative spaces.}{\normalsize \par}

{\normalsize{}}{\normalsize \par}

$\phantom{}$

\textbf{Keywords:} Noncommutative graphene ; Dirac equation ; dynamical
noncommutative space ; zero temperature limit
\end{abstract}

\keywords{Noncommutative graphene ; Dirac equation ; dynamical noncommutative
space ; zero temperature limit}

\maketitle

\section{{\normalsize{}Intoduction}}

Graphene is a crystalline two-dimensional (2D) material, which is
an allotrope of carbon consisting of a single layer of atoms arranged
in a 2D honeycomb lattice \cite{key-1,key-2}. Graphene is thought
about to be the world's thinnest, strongest and most conductive material
of both electricity and heat. Its charge carriers exhibit giant intrinsic
mobility, have zero effective mass, and can travel for micrometers
without scattering at room temperature \cite{key-3}. It is considered
to have the potential to revolutionize whole industries in the fields
of electricity, conductivity, energy generation, sensors and more.
So there is strong motivation for researchers around the world to
study the properties of graphene (see e.g. \cite{key-4,key-5,key-6,key-7,key-8}).
 It is also well-known that graphene plays a substantial role in
various branches of science. Its experimental realization has opened
new horizons in material science and condensed-matter physics. Graphene
is considered as one of the most famous materials in the history in
electronics and material science for its outstanding several unique
mechanical, optical, electrical, transport and thermodynamic properties
\cite{key-9,key-10,key-11,key-12,key-13,key-14,key-15,key-16}. 

Moreover, graphene physics is one of rich and vibrant fields of investigation,
which has attracted a lot of attention from scientists since experimental
observations revealed the existence of electrical charge carriers
behaving like massless Dirac quasi-particles \cite{key-1,key-2,key-17,key-18}.
It was observed that the low-energy electronic excitations at the
corners of graphene Brillouin zone can be described by a 2D Dirac
fermions with linear dispersion relation \cite{key-9,key-18}. Later,
this effect led to testing many aspects of relativistic phenomena
that usually requires large energy in experiments \cite{key-19,key-20,key-21}.

On the other side, the investigation of quantum systems in a noncommutative
(NC) space has been a matter of much interest in the recent years.
Many types of noncommutativity have been considered (see \cite{key-22,key-23,key-24,key-25,key-26,key-27}
for an overview).  However, one type of noncommutativity of space,
is particularly important to us, it is called dynamical noncommutativity
(position-dependent noncommutativity), characterized by $\Theta$
being considered as a function of coordinates, i.e. $\Theta\text{\textrightarrow}\Theta(X,Y)$.

Clearly, studying NC geometry is very important for understanding
phenomena at short distances and has great impact in many areas of
modern physics such as quantum physics, high energy, cosmology, gravity.
For a review of NC quantum mechanics and NC field theories, see \cite{key-28,key-29,key-30,key-31,key-32,key-33,key-34,key-35,key-36,key-37,key-38}. 

In this article, we attempt to examine the effects of dynamical noncommutative
(DNC) space on the graphene in the presence of an external constant
magnetic field. In the same context, Bastos et al \cite{key-39} have
studied graphene in the framework of NC quantum mechanics and determined
the Hamiltonian and the corresponding energy spectrum. Likewise,
Santos et al \cite{key-40} employed the statistics theory to investigate
the thermodynamical properties of graphene in a NC phase-space in
the presence of a constant magnetic field. They exactly found the
main thermodynamical properties of graphene in NC phase-space. In
addition, Boumali \cite{key-41} calculated the thermal properties
of graphene under a magnetic field through the 2D Dirac oscillator
(DO). He showed that using the approach of effective mass, the model
of a 2D DO can be used to describe the thermal properties of graphene
under an uniform magnetic field. The main thermodynamic quantities
of graphene have been found by using an approach based on the zeta
function. More recently, Khordad et al \cite{key-42}  considered
a NC description of graphene and employed extensive and non-extensive
entropies to study magnetic susceptibility of graphene in NC phase-space...
etc.

$\phantom{}$

This paper is outlined as follows. In Section. \ref{II}, the DNC
space is shortly reviewed. In Section. \ref{III}, the 2D graphene
is investigated, where we briefly review the graphene. In sub-section
\ref{III.A}, we extend the problem to DNC space. Then, in sub-sections
\ref{III.B} \& \ref{C} we obtain the solution of Dirac equation
and its spectrum in DNC space. In sub-section \ref{III.D}, based
on the second-order correction on energy, an upper bound on the DNC
parameter is found. In sub-section \ref{III.E}, the thermodynamics
properties of graphene at zero temperature are investigated. We present
our conclusion in Section. \ref{IV}.

\section{\label{II}dynamical noncommutativive space}

Let us first review the substantial relations of the DNC space algebra.
It is well-known that in NC spaces (at the tiny scale), the position
coordinates do not commute with each other anymore. The NC coordinates
satisfy the following deformed commutation relation:
\begin{equation}
\left[x_{\mu}^{nc},x_{\nu}^{nc}\right]=i\Theta_{\mu\nu},\label{eq:1-3}
\end{equation}
where $\Theta_{\mu\nu}$ is an anti-symmetric tensor. The simplest
case is the situation where $\Theta$ is constant, which we call non-dynamical
noncommutative space (NC space or $\Theta$-space).

In Ref.\cite{key-23}, an interesting generalization to DNC spaces
was proposed so that the fundamental objects in this type of DNC spaces
are string like therefore it is a good motivation to study physics
in these spaces. In this reference, $\Theta_{\mu\nu}$ is chosen to
be a function of coordinates as $\theta(X,Y)=\Theta\left(1+\tau Y^{2}\right)$.
Certainly, there is a multitude of other possibilities, such as $\theta(X,Y)=\Theta/[1+\Theta\alpha\left(1+Y^{2}\right)]$
(following Gomes et al in \cite{key-43}). 

The commutation relations for a 2D DNC space (or $\tau$-space) are
as follows \cite{key-23}: 
\begin{equation}
\begin{array}{cccc}
\left[X,Y\right]= & i\Theta\left(1+\tau Y^{2}\right), & \left[Y,P_{y}\right]= & i\hbar\left(1+\tau Y^{2}\right),\\
\left[X,P_{x}\right]= & i\hbar\left(1+\tau Y^{2}\right), & \left[Y,P_{x}\right]= & 0,\\
\left[X,P_{y}\right]= & 2i\tau Y\left(\Theta P_{y}+\hbar X\right), & \left[P_{x},P_{y}\right]= & 0.
\end{array}\label{eq:2-2}
\end{equation}

It is worth mentioning that $\tau$ and $\Theta$ have dimensions
of $\text{L}{}^{\text{\textminus}1}$ and $\text{L}^{2}$, respectively.

In the limit $\tau\rightarrow0$, we recover the non-dynamical ($\Theta$-noncommutative)
commutation relations: 
\begin{equation}
\begin{array}{cccc}
\left[x^{nc},y^{nc}\right]= & i\Theta, & \left[y^{nc},p_{y}^{nc}\right]= & i\hbar,\\
\left[x^{nc},p_{x}^{nc}\right]= & i\hbar, & \left[y^{nc},p_{x}^{nc}\right]= & 0,\\
\left[x^{nc},p_{y}^{nc}\right]= & 0, & \left[p_{x}^{nc},p_{y}^{nc}\right]= & 0.
\end{array}\label{eq:3}
\end{equation}

The coordinate $X$ and the momentum $P_{y}$ are not Hermitian, which
make the Hamiltonian that depends on these variables be non-Hermitian.
We may represent algebra (\ref{eq:2-2}) in terms of the standard
Hermitian NC variables operators $x^{nc},y^{nc},p_{x}^{nc},p_{y}^{nc}$
as
\begin{equation}
\begin{array}{cc}
X=\left(1+\tau\left(y^{nc}\right)^{2}\right)x^{nc},\; & Y=y^{nc},\\
P_{y}=\left(1+\tau\left(y^{nc}\right)^{2}\right)p_{y}^{nc},\; & P_{x}=p_{x}^{nc}.
\end{array}\label{eq:4-3}
\end{equation}

From this representation, we can see that some of the operators involved
above are no longer Hermitian. However, Fring et al \cite{key-23}
 has fixed this problem by converting the non-Hermitian variables
into a Hermitian one. Using a Dyson map $\eta O\eta^{-1}=o=O^{\dagger}$
(with $\eta=(1+\tau Y^{2})^{-\frac{1}{2}}$), so the new Hermitian
variables $x$, $y$, $p_{x}$ and $p_{y}$ in terms of NC variables
are expressed as follows 
\begin{equation}
\begin{array}{cccc}
x=\eta X\eta^{-1} & = & (1+\tau Y^{2})^{-\frac{1}{2}}X(1+\tau Y^{2})^{\frac{1}{2}}\quad\quad\quad\\
 & = & (1+\tau\left(y^{nc}\right)^{2})^{\frac{1}{2}}x^{nc}(1+\tau\left(y^{nc}\right)^{2})^{\frac{1}{2}},\\
y=\eta Y\eta^{-1} & = & (1+\tau\left(y^{nc}\right)^{2})^{-\frac{1}{2}}y^{nc}(1+\tau\left(y^{nc}\right)^{2})^{\frac{1}{2}} & =y^{nc},\\
p_{x}=\eta P_{x}\eta^{-1} & = & (1+\tau\left(y^{nc}\right)^{2})^{-\frac{1}{2}}p_{x}^{nc}(1+\tau\left(y^{nc}\right)^{2})^{\frac{1}{2}} & =p_{x}^{nc},\\
p_{y}=\eta P_{y}\eta^{-1} & = & (1+\tau\left(y^{nc}\right)^{2})^{-\frac{1}{2}}P_{y}(1+\tau\left(y^{nc}\right)^{2})^{\frac{1}{2}}\\
 & = & (1+\tau\left(y^{nc}\right)^{2})^{\frac{1}{2}}p_{y}^{nc}(1+\tau\left(y^{nc}\right)^{2})^{\frac{1}{2}}.
\end{array}\label{eq:4-1}
\end{equation}

The new DNC variables satisfy the following commutation relations
\begin{equation}
\begin{array}{cccc}
\left[x,y\right]= & i\Theta\left(1+\tau y^{2}\right), & \left[y,p_{y}\right]= & i\hbar\left(1+\tau y^{2}\right),\\
\left[x,p_{x}\right]= & i\hbar\left(1+\tau y^{2}\right), & \left[y,p_{x}\right]= & 0,\\
\left[x,p_{y}\right]= & 2i\tau y\left(\Theta p_{y}+\hbar x\right), & \left[p_{x},p_{y}\right]= & 0.
\end{array}\label{eq:6-2}
\end{equation}

Now, through Bopp-shift transformation, one can express the NC variables
in terms of the standard commutative variables \cite{key-31}: 
\begin{equation}
\begin{array}{cc}
x^{nc}=x^{s}-\frac{\Theta}{2\hbar}p_{y}^{s},\: & p_{x}^{nc}=p_{x}^{s},\\
y^{nc}=y^{s}+\frac{\Theta}{2\hbar}p_{y}^{s},\: & p_{y}^{nc}=p_{y}^{s},
\end{array}\label{eq:7-1}
\end{equation}
where the index $s$ refers to the standard commutative space. Noting
that in the DNC space, there is a minimum length for $X$ in a simultaneous
$X$, $Y$ measurement \cite{key-23}:
\begin{equation}
\triangle X_{\text{min}}=\Theta\sqrt{\tau}\sqrt{1+\tau\left\langle Y\right\rangle _{\rho}^{2}},\label{eq:7-2}
\end{equation}
with no minimal length in $Y$. As well, in a simultaneous $Y$, $P_{y}$
measurement, we have a minimal momentum as
\begin{equation}
\triangle\left(P_{y}\right)_{\text{min}}=\hbar\sqrt{\tau}\sqrt{1+\tau\left\langle Y\right\rangle _{\rho}^{2}}.\label{eq:7-3}
\end{equation}

As mentioned before, the motivation behind dynamical noncommutativity
is that objects in 2D space are string-like \cite{key-23}.

\section{{\normalsize{}\label{III}DYNAMICAL NONCOMMUTATIVE GRAPHENE}}

Graphene is a 2D stable allotrope of carbon. It used for constructing
other nanoscale carbons, so that it is the basic structural element
of the other carbon allotropes such as 3D Graphite, 1D carbon nanotube
(CNT), 0D fullerene, i.e. C60, C50, C6 and 3D Diamond. 

The time-independent Dirac equation simply reads
\begin{equation}
H_{D}\psi(r)=E\psi(r),\label{eq:8-1}
\end{equation}
the Dirac Hamiltonian is given by
\begin{equation}
H_{D}=c\overrightarrow{\alpha}.\overrightarrow{p}+\beta mc^{2},\label{eq:9}
\end{equation}
$\alpha$, $\beta$ are the Dirac matrices. In the case of graphene,
we have masseless particles move through the honeycomb lattice where
$c\longrightarrow v_{F}\simeq10^{6}m.s^{-1}$(Fermi velocity) \cite{key-6},
which gives
\begin{equation}
H_{D}=v_{F}\overrightarrow{\sigma}.\overrightarrow{p}.\label{eq:10-2}
\end{equation}

From electronics peoperties point of view, graphene is a zero-gap
semiconductor, in which low-energy quasiparticles within each valley
can formally be described by the Dirac Hamiltonian. The wave function
$\psi(r)$ gives the electron states around the points $K_{i}$ and
the Dirac block diagonalized Hamiltonian is given by
\begin{equation}
\left.H_{D}\right|_{K_{i}}=\left(\begin{array}{cccc}
H_{k_{1}} & 0 & 0 & \cdots\\
0 & H_{k_{2}} & 0 & 0\\
0 & 0 & H_{k_{3}} & 0\\
\vdots & 0 & 0 & \ddots
\end{array}\right).\label{eq:xx}
\end{equation}

The most important blocks amid the blocks of the Hamiltonian (\ref{eq:xx})
are the blocks of the two specific wavenumbers, namely the Dirac points
$K$ and $K^{'}$ respectively, specified by \cite{key-44} 
\begin{equation}
K=\frac{2\pi}{3}\left(\frac{1}{\sqrt{3}}\right),\quad K^{'}=\frac{2\pi}{3}\left(\frac{-1}{\sqrt{3}}\right).\label{eq:12-1}
\end{equation}

These two Dirac points control the elementary excitation of graphene.
The action of electrons around the Dirac points $K$ and $K^{'}$
at the corners of Brillouin zone is descibed by the Dirac Hamiltonian
where in the case of $K^{'}$, we have $\overrightarrow{\sigma}^{\ast}=\left(\sigma_{x},-\sigma_{y},\sigma_{z}\right)$.
In this manner, we can write 
\begin{equation}
\left.H_{D}\right|_{K,K^{'}}=\left(\begin{array}{cc}
H_{K} & 0\\
0 & H_{K^{'}}
\end{array}\right)=v_{F}\left(\begin{array}{cc}
\overrightarrow{\sigma}.\overrightarrow{p} & 0\\
0 & \overrightarrow{\sigma}^{\ast}.\overrightarrow{p}
\end{array}\right),\label{eq:13-1}
\end{equation}
consequently $\psi_{K,K^{'}}(r)$ is the two-component wavefunction,
which gives the electron states around the the Dirac points $K$ and
$K^{'}$.

\subsection{\label{III.A}EXTENSION TO DYNAMICAL NONCOMMUTATIVE SPACE}

Let us consider a layer of graphene in an external constant magnetic
field along the $z$ axis. We introduce $B\hat{z}$ through the minimal
coupling to the vector potential, so we define the canonical momentum
as
\begin{equation}
\overrightarrow{p}^{s}\longrightarrow\overrightarrow{p}^{s}-\frac{e}{c}\overrightarrow{A}^{s},\label{eq:14}
\end{equation}
where $\overrightarrow{A}^{s}$ is the electomagnetic vector potential
and in the symmetric gauge is given by
\begin{equation}
\overrightarrow{A}^{s}=\frac{B}{2}(-y^{s},x^{s},0).\label{eq:15}
\end{equation}

The 2d time-independent Dirac equation around two Dirac points $K$
and $K'$ reads 
\begin{equation}
H_{D}\psi_{K,K^{'}}(r^{s})=v_{F}\left(\begin{array}{cc}
\overrightarrow{\sigma}.\left(\overrightarrow{p}^{s}-\frac{e}{c}\overrightarrow{A}^{s}\right) & 0\\
0 & \overrightarrow{\sigma}^{\ast}.\left(\overrightarrow{p}^{s}-\frac{e}{c}\overrightarrow{A}^{s}\right)
\end{array}\right)\psi_{K,K^{'}}(r^{s})=E_{K,K^{'}}\psi_{K,K^{'}}(r^{s}).\label{eq:16}
\end{equation}

The Dirac equation around Dirac point $K$ is given by

\begin{equation}
H_{K}\psi^{K}(r^{s})=v_{F}\overrightarrow{\sigma}.\left(\overrightarrow{p}^{s}-\frac{e}{c}\overrightarrow{A}^{s}\right)\psi^{K}(r^{s})=E_{K}\psi^{K}(r^{s}),\label{eq:17}
\end{equation}
where $\psi^{K}=\left(\begin{array}{cc}
\phi^{A} & \phi^{B}\end{array}\right)^{t}$ is two dimensional eigenstante, i.e. an eigenvector that describes
the probablility of an electron state to be on sub-lattice A in the
upper component, or on the sub-lattice B in the lower component of
the eigenstate, and $t$ denotes transpose.  A similar equation can
also be obtained for the Dirac point $K'$ with $\psi^{K^{'}}=\left(\begin{array}{cc}
\phi^{A'} & \phi^{B'}\end{array}\right)^{t}$.

$\overrightarrow{\sigma}$ are the three 2\texttimes 2 Pauli matrices,
which are given by

\begin{equation}
\sigma_{x}=\alpha_{1}=\left(\begin{array}{cc}
0 & 1\\
1 & 0
\end{array}\right),\:\sigma_{y}=\alpha_{2}=\left(\begin{array}{cc}
0 & -i\\
i & 0
\end{array}\right)\text{ and }\sigma_{z}=\beta=\left(\begin{array}{cc}
1 & 0\\
0 & -1
\end{array}\right).\label{eq:18}
\end{equation}

Therefore, for the Dirac point $K$, one has the following Hamiltonian
\begin{equation}
H_{K}\left(x_{i}^{s},p_{i}^{s}\right)=v_{F}\left\{ \alpha_{1}p_{x}^{s}+\alpha_{2}p_{y}^{s}+\frac{\hbar}{2l_{B}^{2}}\left(\alpha_{1}y^{s}-\alpha_{2}x^{s}\right)\right\} ,\label{eq:19}
\end{equation}
where we have used $l_{B}=\sqrt{\frac{c\hbar}{eB}}$, which is the
magnetic length.

The above Hamiltonian in DNC space turns to
\begin{equation}
H_{K}\left(x_{i},p_{i}\right)=v_{F}\left\{ \alpha_{1}p_{x}+\alpha_{2}p_{y}+\frac{\hbar}{2l_{B}^{2}}\left(\alpha_{1}y-\alpha_{2}x\right)\right\} .\label{eq:8}
\end{equation}

Now, using equation (\ref{eq:4-1}), we express the Hamiltonian above
in terms of NC variables
\begin{equation}
\begin{array}{c}
H_{K}\left(x_{i}^{nc},p_{i}^{nc}\right)=v_{F}\alpha_{1}p_{x}^{nc}+v_{F}\alpha_{2}\left(1+\tau\left(y^{nc}\right)^{2}\right)^{\frac{1}{2}}p_{y}^{nc}\left(1+\tau\left(y^{nc}\right)^{2}\right)^{\frac{1}{2}}\\
+v_{F}\frac{\hbar}{2l_{B}^{2}}\left[\alpha_{1}y^{nc}-\alpha_{2}\left(1+\tau\left(y^{nc}\right)^{2}\right)^{\frac{1}{2}}x^{nc}\left(1+\tau\left(y^{nc}\right)^{2}\right)^{\frac{1}{2}}\right].
\end{array}\label{eq:9-1}
\end{equation}

Since $\tau$ is very small, the parentheses can be expanded to the
first order using 
\begin{equation}
\left(1+\tau\left(y^{nc}\right)^{2}\right)^{\frac{1}{2}}=1+\frac{1}{2}\tau\left(y^{nc}\right)^{2},\label{eq:10}
\end{equation}
thus, equation (\ref{eq:9-1}) becomes 
\begin{equation}
\begin{array}{c}
H_{K}\left(x_{i}^{nc},p_{i}^{nc}\right)=v_{F}\left[\alpha_{1}p_{x}^{nc}+\alpha_{2}\left\{ p_{y}^{nc}+\frac{1}{2}\tau\left(y^{nc}\right)^{2}p_{y}^{nc}+\frac{1}{2}\tau p_{y}^{nc}\left(y^{nc}\right)^{2}\right\} \right]\\
+v_{F}\frac{\hbar}{2l_{B}^{2}}\left[\alpha_{1}y^{nc}-\alpha_{2}\left\{ x^{nc}+\frac{1}{2}\tau\left(y^{nc}\right)^{2}x^{nc}+\frac{1}{2}\tau x^{nc}\left(y^{nc}\right)^{2}\right\} \right].
\end{array}\label{eq:10-1}
\end{equation}

Now using the Bopp-shift transformation (\ref{eq:7-1}), the Hamiltonian
(\ref{eq:10-1}) can be expressed in terms of the standard commutative
variables

\begin{equation}
\begin{array}{c}
H_{K}\left(x_{i}^{s},p_{i}^{s}\right)=v_{F}\alpha_{1}p_{x}^{s}+v_{F}\alpha_{2}p_{y}^{s}+v_{F}\alpha_{2}\left\{ \frac{1}{2}\tau\left(y^{s}+\frac{\Theta}{2\hbar}p_{x}^{s}\right)^{2}p_{y}^{s}+\frac{1}{2}\tau p_{y}^{s}\left(y^{s}+\frac{\Theta}{2\hbar}p_{x}^{s}\right)^{2}\right\} \\
+v_{F}\frac{\hbar}{2l_{B}^{2}}\left[\alpha_{1}\left(y^{s}+\frac{\Theta}{2\hbar}p_{x}^{s}\right)-\alpha_{2}\left\{ x^{s}-\frac{\Theta}{2\hbar}p_{y}^{s}+\frac{\tau}{2}\left(y^{s}+\frac{\Theta}{2\hbar}p_{x}^{s}\right)^{2}\left(x^{s}-\frac{\Theta}{2\hbar}p_{y}^{s}\right)+\frac{1}{2}\tau\left(x^{s}-\frac{\Theta}{2\hbar}p_{y}^{s}\right)\left(y^{s}+\frac{\Theta}{2\hbar}p_{x}^{s}\right)^{2}\right\} \right].
\end{array}\label{eq:12}
\end{equation}

Therefore, to the first order in $\Theta$ and $\tau$, we have (the
terms containing $\Theta\tau$ are also neglected) 
\begin{equation}
\begin{array}{c}
H_{K}\left(x_{i}^{s},p_{i}^{s}\right)=v_{F}\left[\alpha_{1}p_{x}^{s}+\alpha_{2}\left\{ p_{y}^{s}+\frac{\tau}{2}\left(y^{s}\right)^{2}p_{y}^{s}+\frac{\tau}{2}p_{y}^{s}\left(y^{s}\right)^{2}\right\} \right]\\
+v_{F}\frac{\hbar}{2l_{B}^{2}}\left[\alpha_{1}\left(y^{s}+\frac{\Theta}{2\hbar}p_{x}^{s}\right)-\alpha_{2}\left\{ x^{s}-\frac{\Theta}{2\hbar}p_{y}^{s}+\tau x^{s}\left(y^{s}\right)^{2}\right\} \right],
\end{array}\label{eq:13}
\end{equation}
which may be written as 
\begin{equation}
H_{K}=H_{K}^{(0)}+H_{K}^{(\Theta)}+H_{K}^{(\tau)},\label{eq:26}
\end{equation}
where
\begin{equation}
H_{K}^{(0)}=v_{F}\left\{ \alpha_{1}p_{x}^{s}+\alpha_{2}p_{y}^{s}+\frac{\hbar}{2l_{B}^{2}}\left(\alpha_{1}y^{s}-\alpha_{2}x^{s}\right)\right\} ,\label{eq:26-1}
\end{equation}
\begin{equation}
H_{K}^{(\Theta)}=v_{F}\frac{\Theta}{4l_{B}^{2}}\left(\alpha_{1}p_{x}^{s}+\alpha_{2}p_{y}^{s}\right),\label{eq:27}
\end{equation}
\begin{equation}
H_{K}^{(\tau)}=v_{F}\frac{\tau}{2}\alpha_{2}\left\{ \left(y^{s}\right)^{2}p_{y}^{s}+p_{y}^{s}\left(y^{s}\right)^{2}-\frac{\hbar}{l_{B}^{2}}x^{s}\left(y^{s}\right)^{2}\right\} .\label{eq:28}
\end{equation}

A similar set of equations can also be obtained for the Dirac point
$K^{'}$.

\subsection{\label{III.B}Unperturbed System }

\selectlanguage{french}%
Using equation (\ref{eq:18}), the Dirac Hamiltonian in commutative
space (\ref{eq:26-1}) becomes 
\begin{equation}
\begin{array}{ccc}
H_{K}^{(0)} & = & v_{F}\left\{ \sigma_{x}p_{x}^{s}+\sigma_{y}p_{y}^{s}-e\sigma_{x}A_{x}^{s}-e\sigma_{y}A_{y}^{s}\right\} \\
 & = & v_{F}\left\{ \left(\begin{array}{cc}
0 & 1\\
1 & 0
\end{array}\right)p_{x}^{s}+\left(\begin{array}{cc}
0 & -i\\
i & 0
\end{array}\right)p_{y}^{s}+\frac{\hbar}{2l_{B}^{2}}\left(\begin{array}{cc}
0 & 1\\
1 & 0
\end{array}\right)y^{s}-\frac{\hbar}{2l_{B}^{2}}\left(\begin{array}{cc}
0 & -i\\
i & 0
\end{array}\right)x^{s}\right\} ,
\end{array}\label{eq:123}
\end{equation}
or in more compact form 
\begin{equation}
H_{K}^{(0)}=v_{F}\left(\begin{array}{cc}
0 & p_{x}^{s}-ip_{y}^{s}+\frac{\hbar}{2l_{B}^{2}}\left(y^{s}+ix^{s}\right)\\
p_{x}^{s}+ip_{y}^{s}+\frac{\hbar}{2l_{B}^{2}}\left(y^{s}-ix^{s}\right) & 0
\end{array}\right).\label{eq:124}
\end{equation}

By introducing $\kappa=\frac{\hbar}{2l_{B}^{2}}$, we have 
\begin{equation}
H_{K}^{(0)}=v_{F}\left(\begin{array}{cc}
0 & p_{x}^{s}-ip_{y}^{s}+\kappa\left(y^{s}+ix^{s}\right)\\
p_{x}^{s}+ip_{y}^{s}+\kappa\left(y^{s}-ix^{s}\right) & 0
\end{array}\right)=v_{F}\left(\begin{array}{cc}
0 & h_{12}\\
h_{21} & 0
\end{array}\right).\label{eq:35}
\end{equation}

Let us now introduce the following creation and annihilation operators
\begin{equation}
a_{x}=\frac{1}{\sqrt{2\kappa\hbar}}\left(\kappa x^{s}+ip_{x}^{s}\right),\text{ }a_{x}^{\dagger}=\frac{1}{\sqrt{2\alpha\hbar}}\left(\kappa x^{s}-ip_{x}^{s}\right),\label{eq:36}
\end{equation}
\begin{equation}
a_{y}=\frac{1}{\sqrt{2\kappa\hbar}}\left(\alpha y^{s}+ip_{y}^{s}\right),\text{ }a_{y}^{\dagger}=\frac{1}{\sqrt{2\kappa\hbar}}\left(\alpha y^{s}-ip_{y}^{s}\right),\label{eq:37}
\end{equation}
which satisfy the following commutations relations 
\begin{equation}
\left[a_{i},a_{i}^{\dagger}\right]=1,\,\left[a_{i},a_{i}\right]=\left[a_{i}^{\dagger},a_{i}^{\dagger}\right]=0,\;\:i=x,y.\label{eq:38}
\end{equation}

So different components of the Hamiltonian could be written in terms
of the creation and annihilation operators as follows 
\begin{equation}
h_{12}=\sqrt{2\kappa\hbar}\left(a_{y}^{\dagger}+ia_{x}^{\dagger}\right),\label{eq:39}
\end{equation}
\begin{equation}
h_{21}=h_{12}^{\dagger}=\sqrt{2\kappa\hbar}\left(a_{y}-ia_{x}\right).\label{eq:40}
\end{equation}

\selectlanguage{english}%
Thus, the Hamiltonian (\ref{eq:35}) takes the following form 
\begin{equation}
H_{K}^{(0)}=v_{F}\left(\begin{array}{cc}
0 & 2\sqrt{\frac{\kappa\hbar}{2}}\left(a_{y}^{\dagger}+ia_{x}^{\dagger}\right)\\
2\sqrt{\frac{\kappa\hbar}{2}}\left(a_{y}-ia_{x}\right) & 0
\end{array}\right)=v_{F}\left(\begin{array}{cc}
0 & \frac{g}{\sqrt{2}}\left(a_{y}^{\dagger}+ia_{x}^{\dagger}\right)\\
\frac{g}{\sqrt{2}}\left(a_{y}-ia_{x}\right) & 0
\end{array}\right),\label{eq:41}
\end{equation}
in which the parameter $g=2\sqrt{\kappa\hbar}$ describes the coupling
between different states in commutative space. 

\selectlanguage{french}%
Let us return to the Dirac equation
\begin{equation}
H_{K}^{(0)}\left|\psi_{(0)}^{K}\right\rangle =E_{K}^{(0)}\left|\psi_{(0)}^{K}\right\rangle ,\label{eq:136-1}
\end{equation}
\foreignlanguage{english}{where $E_{K}^{(0)}$, $\left|\psi_{(0)}^{K}\right\rangle $
are the eigenvalues and eigenkets of the Dirac Hamiltonian in commutative
space, respectively. In two dimensions, $\left|\psi_{(0)}^{K}\right\rangle $
is written as} 
\begin{equation}
\left|\psi_{(0)}^{K}\right\rangle =\left(\begin{array}{c}
\left|\psi_{A}\right\rangle \\
\left|\psi_{B}\right\rangle 
\end{array}\right).\label{eq:137}
\end{equation}

By setting $C=\frac{1}{\sqrt{2}}\left(a_{y}-ia_{x}\right)$, and inserting
equation (\ref{eq:41}) in equation (\ref{eq:136-1}), we obtain the
following system of equations 
\begin{equation}
v_{F}\left(\begin{array}{cc}
0 & gC^{\dagger}\\
gC & 0
\end{array}\right)\left(\begin{array}{c}
\left|\psi_{A}\right\rangle \\
\left|\psi_{B}\right\rangle 
\end{array}\right)=E_{K}^{(0)}\left(\begin{array}{c}
\left|\psi_{A}\right\rangle \\
\left|\psi_{B}\right\rangle 
\end{array}\right),\label{eq:138-1}
\end{equation}
\begin{equation}
-E_{K}^{(0)}\left|\psi_{A}\right\rangle +v_{F}gC^{\dagger}\left|\psi_{B}\right\rangle =0,\label{eq:139-1}
\end{equation}
\begin{equation}
v_{F}gC\left|\psi_{A}\right\rangle -E_{K}^{(0)}\left|\psi_{B}\right\rangle =0.\label{eq:140-1}
\end{equation}

Equations (\ref{eq:139-1}) and (\ref{eq:140-1}) give
\begin{equation}
\left|\psi_{B}\right\rangle =\frac{v_{F}gC}{E_{K}^{(0)}}\left|\psi_{A}\right\rangle ,\label{eq:141-1}
\end{equation}
\begin{equation}
-\left(E_{K}^{(0)}\right)^{2}\left|\psi_{A}\right\rangle +v_{F}^{2}g^{2}C^{\dagger}C\left|\psi_{B}\right\rangle =0,\label{eq:47-1}
\end{equation}
so
\begin{equation}
\left(v_{F}^{2}g^{2}C^{\dagger}C-\left(E_{K}^{(0)}\right)^{2}\right)\left|\psi_{A}\right\rangle =0.\label{eq:47-2}
\end{equation}

\selectlanguage{english}%
In the basis of $C^{\dagger}C=N$, we have \foreignlanguage{french}{
\begin{equation}
\left[v_{F}^{2}g^{2}N-\left(E_{K}^{(0)}\right)^{2}\right]\left|\psi_{A}\right\rangle =0,\,\mbox{with}\;N\left|\psi_{A}\right\rangle =n\left|\psi_{A}\right\rangle .\label{eq:143-1}
\end{equation}
}

Thus, the energy spectrum is given by\foreignlanguage{french}{
\begin{equation}
\left.E_{n}^{\pm}\right|_{K}^{(0)}=\pm\sqrt{v_{F}^{2}g^{2}n},\label{eq:144-1}
\end{equation}
}which can be rewritten as \foreignlanguage{french}{
\begin{equation}
\left.E_{n}^{\pm}\right|_{K}^{(0)}=\pm v_{F}\frac{\hbar}{l_{B}}\sqrt{2n},\,n=0,1,2,...\label{eq:145-1}
\end{equation}
}

\selectlanguage{french}%
The result is in good agreement with that of ordinary quantum mechanics
\cite{key-39,key-40}.\foreignlanguage{english}{ Now, if we consider
equations (\ref{eq:139-1}, \ref{eq:140-1}), and substitute the equation
(\ref{eq:145-1}), we get the corresponding wave function for the
Dirac point $K$}

\begin{equation}
\left|\psi_{k}^{(0)}\right\rangle =\left(\begin{array}{cc}
\left|\psi_{A}\right\rangle  & \pm i\left|\psi_{B}\right\rangle \end{array}\right)^{t}.\label{eq:146-1}
\end{equation}

Using the same method, the eigenvalues and eigenstates of the Dirac
point \foreignlanguage{english}{$K^{'}$} can be obtained.

\selectlanguage{english}%

\subsection{\label{C}Perturbed System }

If the dynamical noncommutativity parameter $\tau$ is non-zero, should
be very small compared to the energy scales of the system, one can
always treat the DNC effects as some perturbations of the commutative
analogue, so we use time-independent perturbation theory to study
the system in $\tau$- space.

if it is nonzero, should be very small

From Equations (\ref{eq:36}) and (\ref{eq:37}) we have
\begin{equation}
\begin{array}{c}
x^{s}=\frac{1}{2\Gamma}\left(a_{d}+a_{d}^{\dagger}+a_{g}+a_{g}^{\dagger}\right),\:\text{ }y^{s}=\frac{i}{2\Gamma}\left(a_{d}-a_{d}^{\dagger}-a_{g}+a_{g}^{\dagger}\right),\\
p_{x}^{s}=\frac{1}{2\Gamma}\left(-a_{d}+a_{d}^{\dagger}-a_{g}+a_{g}^{\dagger}\right),\:p_{y}^{s}=\frac{i}{2\Gamma}\left(a_{d}+a_{d}^{\dagger}-a_{g}-a_{g}^{\dagger}\right),
\end{array}\label{eq:46-1}
\end{equation}
with
\begin{equation}
\Gamma=\frac{1}{\sqrt{2}l_{B}}.\label{eq:54-1}
\end{equation}

By introducing
\begin{equation}
n=n_{d}+n_{g}\text{ and }m=n_{d}-n_{g},\label{eq:54-2}
\end{equation}
the eigenkets of the Hamiltonian could be represented as follows
\begin{equation}
\left|n_{d}=\frac{n+m}{2},\:n_{g}=\frac{n-m}{2}\right\rangle ,\label{eq:54-3}
\end{equation}
we distinguish the following states: 

\begin{center}
\begin{tabular}{|c|c|c|}
\hline 
Ground state  & $\qquad$ n=0, m=0 $\qquad$  & $\left|0,0\right\rangle $\tabularnewline
\hline 
\hline 
\multirow{2}{*}{First excited state } & n=1, m=1  & $\left|1,0\right\rangle $\tabularnewline
\cline{2-3} 
 & n=1, m=-1  & $\left|0,1\right\rangle $\tabularnewline
\hline 
\multirow{3}{*}{Second excited state } & n=2, m=2  & $\left|2,0\right\rangle $\tabularnewline
\cline{2-3} 
 & n=2, m=0  & $\left|1,1\right\rangle $\tabularnewline
\cline{2-3} 
 & n=2, m=-2  & $\left|0,2\right\rangle $\tabularnewline
\hline 
\multirow{4}{*}{Third excited state } & n=3, m=3  & $\left|3,0\right\rangle $\tabularnewline
\cline{2-3} 
 & n=3, m=1  & $\left|2,1\right\rangle $\tabularnewline
\cline{2-3} 
 & n=3, m=-1  & $\left|1,2\right\rangle $\tabularnewline
\cline{2-3} 
 & n=3, m=-3 & $\left|0,3\right\rangle $\tabularnewline
\hline 
\end{tabular}
\par\end{center}

Now, let move to calculate the different elements in our perturbed
Hamiltonian. We start with $x^{s}y^{s2}$, thus using equation (\ref{eq:46-1})
we have
\begin{equation}
\begin{array}{c}
x^{s}y^{s2}=\frac{-1}{8\Gamma^{3}}\left[a_{d}^{3}-a_{d}^{2}a_{d}^{\dagger}-a_{d}^{2}a_{g}-a_{d}^{2}a_{g}^{\dagger}-a_{d}a_{d}^{\dagger}a_{g}^{\dagger}-a_{d}a_{d}^{\dagger}a_{d}+a_{d}a_{d}^{\dagger2}\right.\\
+a_{d}a_{d}^{\dagger}a_{g}-a_{d}a_{g}a_{d}+a_{d}a_{g}a_{d}^{\dagger}+a_{d}a_{g}^{2}-a_{d}a_{g}a_{g}^{\dagger}+a_{d}a_{g}^{\dagger}a_{d}-a_{d}a_{g}^{\dagger}a_{d}^{\dagger}\\
-a_{d}a_{g}^{\dagger}a_{g}+a_{d}a_{g}^{\dagger2}a_{d}^{\dagger}a_{d}^{2}-a_{d}^{\dagger}a_{d}a_{d}^{\dagger}-a_{d}^{\dagger}a_{d}a_{g}+a_{d}^{\dagger}a_{d}a_{g}^{\dagger}-a_{d}^{\dagger2}a_{d}+a_{d}^{\dagger3}+a_{d}^{\dagger2}a_{g}\\
-a_{d}^{\dagger2}a\dagger_{g}-a_{d}^{\dagger}a_{g}a_{d}+a_{d}^{\dagger}a_{g}a_{d}^{\dagger}+a_{d}^{\dagger}a_{g}^{2}-a_{d}^{\dagger}a_{g}a_{g}^{\dagger}+a_{d}^{\dagger}a_{g}^{\dagger}a_{d}-a_{d}^{\dagger}a_{g}^{\dagger}a_{d}^{\dagger}-a_{d}^{\dagger}a_{g}^{\dagger}a_{g}\\
+a_{d}^{\dagger}a_{g}^{\dagger2}+a_{g}a_{d}^{2}-a_{g}a_{d}a_{d}^{\dagger}-a_{g}a_{d}a_{g}+a_{g}a_{d}a_{g}^{\dagger}-a_{g}a_{d}^{\dagger}a_{d}+a_{g}a_{d}^{\dagger2}+a_{g}a_{d}^{\dagger}a_{g}\\
-a_{g}a_{d}^{\dagger}a_{g}^{\dagger}-a_{g}^{2}a_{d}+a_{g}^{2}a_{d}^{\dagger}+a_{g}^{3}-a_{g}^{2}a_{g}^{\dagger}+a_{g}a_{g}^{\dagger}a_{d}-a_{g}a_{g}^{\dagger}a_{d}^{\dagger}-a_{g}a_{g}^{\dagger}a_{g}+a_{g}a_{g}^{\dagger2}\\
+a_{g}^{\dagger}a_{d}^{2}-a_{g}^{\dagger}a_{d}a_{d}^{\dagger}-a_{g}^{\dagger}a_{d}a_{g}+a_{g}^{\dagger}a_{d}a_{g}^{\dagger}-a_{g}^{\dagger}a_{d}^{\dagger}a_{d}+a_{g}^{\dagger}a_{d}^{\dagger2}+a_{g}^{\dagger}a_{d}^{\dagger}a_{g}-a_{g}^{\dagger}a_{d}^{\dagger}a_{g}^{\dagger}\\
\left.-a_{g}^{\dagger}a_{g}a_{d}+a_{g}^{\dagger}a_{g}a_{d}^{\dagger}+a_{g}^{\dagger}a_{g}^{2}-a_{g}^{\dagger}a_{g}a_{g}^{\dagger}+a_{g}^{\dagger2}a_{d}-a_{g}^{\dagger2}a_{d}^{\dagger}-a_{g}^{\dagger2}a_{g}+a_{g}^{\dagger3}\right].
\end{array}\label{eq:50}
\end{equation}

Corrections due to $x^{s}y^{s2}$ on energy spectrum to the first
order is given by 
\begin{equation}
\left\langle n_{d},n_{g}\right|x^{s}y^{s2}\left|n_{d},n_{g}\right\rangle ,\label{eq:51}
\end{equation}
where for the ground state, we have $\left\langle 0,0\right|x^{s}y^{s2}\left|0,0\right\rangle $,
and using equation (\ref{eq:50}), one can check that 
\begin{equation}
\left\langle 0,0\right|x^{s}y^{s2}\left|0,0\right\rangle =0.\label{eq:52}
\end{equation}

The first excited state is two fold degenerate $\left|1,0\right\rangle $
and $\left|0,1\right\rangle $, so we have the following perturbed
matrix
\begin{equation}
\left(\begin{array}{cc}
\left\langle 0,1\right|x^{s}y^{s2}\left|0,1\right\rangle \: & \:\left\langle 0,1\right|x^{s}y^{s2}\left|1,0\right\rangle \\
\left\langle 1,0\right|x^{s}y^{s2}\left|0,1\right\rangle \: & \:\left\langle 1,0\right|x^{s}y^{s2}\left|1,0\right\rangle 
\end{array}\right).\label{eq:53}
\end{equation}

By doing neccessary calculations we can show that the matrix element
$\left\langle 0,1\right|x^{s}y^{s2}\left|0,1\right\rangle $ vanishes
\begin{equation}
\left\langle 0,1\right|x^{s}y^{s2}\left|0,1\right\rangle =0,\label{eq:54}
\end{equation}
with the same calculations we can also check that the contributions
of the rest of the matrix elements in equation (\ref{eq:53}) vanish.

The second excited state is three fold degenerate $\left|2,0\right\rangle $,
$\left|1,1\right\rangle $, $\left|0,2\right\rangle $ and it is given
by $3\times3$ matrix. One can check again that its elements vanish.

In general one can show that to the first order the term $x^{s}y^{s2}$
in equation (\ref{eq:28}) has no correction to the energy of the
system. Now, we calculate the second order contribution of $x^{s}y^{s2}$.
For the ground state we should find the non-zero matrix elements $\left\langle n_{d},n_{g}\right|x^{s}y^{s2}\left|0,0\right\rangle $,
where the non-zero values are as follows:

\{$\left\langle 3,0\right|a_{d}^{\dagger3}\left|0,0\right\rangle $,
$\left\langle 0,3\right|a_{g}^{\dagger3}\left|0,0\right\rangle $,
$-\left\langle 2,1\right|a_{d}^{\dagger}a_{g}^{\dagger}a_{d}^{\dagger}\left|0,0\right\rangle $,
$-\left\langle 1,2\right|a_{g}^{\dagger}a_{d}^{\dagger}a_{g}^{\dagger}\left|0,0\right\rangle $,
$-\left\langle 0,1\right|a_{d}a_{g}^{\dagger}a_{d}^{\dagger}\left|0,0\right\rangle $,
$-\left\langle 0,1\right|a_{d}a_{d}^{\dagger}a_{g}^{\dagger}\left|0,0\right\rangle $,
$-\left\langle 0,1\right|a_{g}^{\dagger}a_{d}a_{d}^{\dagger}\left|0,0\right\rangle $,
$-\left\langle 1,0\right|a_{d}^{\dagger}a_{g}a_{g}^{\dagger}\left|0,0\right\rangle $,
$-\left\langle 1,0\right|a_{g}a_{d}^{\dagger}a_{g}^{\dagger}\left|0,0\right\rangle $,
$-\left\langle 1,0\right|a_{g}a_{g}^{\dagger}a_{d}^{\dagger}\left|0,0\right\rangle $\}.

$\phantom{}$

Now, we consider the two other terms in equation (\ref{eq:28}) i.e.,
$y^{s2}p_{y}^{s}$, $p_{y}^{s}y^{s2}$. Employing the relation
\begin{equation}
\left[p_{y}^{s},y^{s2}\right]=-2i\hbar y^{s},\label{eq:55}
\end{equation}
we rearrange the terms $y^{s2}p_{y}^{s}+p_{y}^{s}y^{s2}$ as 
\begin{equation}
y^{s2}p_{y}^{s}+p_{y}^{s}y^{s2}=2p_{y}^{s}y^{s2}+2i\hbar y^{s}.\label{eq:56}
\end{equation}

Therefore, we calculate the contibution of $p_{y}^{s}y^{s2}$ thus
we have
\begin{equation}
\begin{array}{c}
x^{s}y^{s2}=-\frac{\hbar}{8\Gamma}\left[a_{d}^{3}-a_{d}^{2}a_{d}^{\dagger}-a_{d}^{2}a_{g}-a_{d}^{2}a_{g}^{\dagger}-a_{d}a_{d}^{\dagger}a_{d}+a_{d}a_{d}^{\dagger2}+a_{d}a_{d}^{\dagger}a_{g}\right.\\
-a_{d}a_{d}^{\dagger}a_{g}^{\dagger}-a_{d}a_{g}a_{d}+a_{d}a_{g}a_{d}^{\dagger}+a_{d}a_{g}^{2}-a_{d}a_{g}a_{g}^{\dagger}+a_{d}a_{g}^{\dagger}a_{d}-a_{d}a_{g}^{\dagger}a_{d}^{\dagger}\\
-a_{d}a_{g}^{\dagger}a_{g}+a_{d}a_{g}^{\dagger2}+a_{d}^{\dagger}a_{d}^{2}-a_{d}^{\dagger}a_{d}a_{d}^{\dagger}-a_{d}^{\dagger}a_{d}a_{g}+a_{d}^{\dagger}a_{d}a_{g}^{\dagger}-a_{d}^{\dagger2}a_{d}+a_{d}^{\dagger3}\\
+a_{d}^{\dagger2}a_{g}-a_{d}^{\dagger2}a\dagger_{g}-a_{d}^{\dagger}a_{g}a_{d}+a_{d}^{\dagger}a_{g}a_{d}^{\dagger}+a_{d}^{\dagger}a_{g}^{2}-a_{d}^{\dagger}a_{g}a_{g}^{\dagger}+a_{d}^{\dagger}a_{g}^{\dagger}a_{d}\\
-a_{d}^{\dagger}a_{g}^{\dagger}a_{d}^{\dagger}-a_{d}^{\dagger}a_{g}^{\dagger}a_{g}+a_{d}^{\dagger}a_{g}^{\dagger2}-a_{g}a_{d}^{2}+a_{g}a_{d}a_{d}^{\dagger}+a_{g}a_{d}a_{g}-a_{g}a_{d}a_{g}^{\dagger}\\
+a_{g}a_{d}^{\dagger}a_{d}-a_{g}a_{d}^{\dagger2}-a_{g}a_{d}^{\dagger}a_{g}+a_{g}a_{d}^{\dagger}a_{g}^{\dagger}+a_{g}^{2}a_{d}-a_{g}^{2}a_{d}^{\dagger}-a_{g}^{3}+a_{g}^{2}a_{g}^{\dagger}\\
-a_{g}a_{g}^{\dagger}a_{d}+a_{g}a_{g}^{\dagger}a_{d}^{\dagger}+a_{g}a_{g}^{\dagger}a_{g}-a_{g}a_{g}^{\dagger2}-a_{g}^{\dagger}a_{d}^{2}+a_{g}^{\dagger}a_{d}a_{d}^{\dagger}+a_{g}^{\dagger}a_{d}a_{g}\\
-a_{g}^{\dagger}a_{d}a_{g}^{\dagger}+a_{g}^{\dagger}a_{d}^{\dagger}a_{d}-a_{g}^{\dagger}a_{d}^{\dagger2}-a_{g}^{\dagger}a_{d}^{\dagger}a_{g}+a_{g}^{\dagger}a_{d}^{\dagger}a_{g}^{\dagger}+a_{g}^{\dagger}a_{g}a_{d}\\
\left.-a_{g}^{\dagger}a_{g}a_{d}^{\dagger}-a_{g}^{\dagger}a_{g}^{2}+a_{g}^{\dagger}a_{g}a_{g}^{\dagger}-a_{g}^{\dagger2}a_{d}+a_{g}^{\dagger2}a_{d}^{\dagger}+a_{g}^{\dagger2}a_{g}-a_{g}^{\dagger3}\right].
\end{array}\label{eq:57}
\end{equation}

One can check that the non-vanishing matrix elements of $p_{y}^{s}y^{s2}$
are the same as $x^{s}y^{s2}$, so we proceed to calculate these non-vanishing
matrix elements. We use the following useful relation 
\begin{equation}
\left(a_{d}^{\dagger}\right)^{n_{d}}\left(a_{g}^{\dagger}\right)^{n_{g}}\left|0,0\right\rangle =\sqrt{n_{d}!n_{g}!}\left|n_{d},n_{g}\right\rangle ,\label{eq:58}
\end{equation}
thus we have
\begin{equation}
\left\langle 3,0\right|a_{d}^{\dagger3}\left|0,0\right\rangle =\sqrt{6}\left\langle 3,0\mid3,0\right\rangle =\sqrt{6},\label{eq:59}
\end{equation}
\begin{equation}
\left\langle 0,3\right|a_{g}^{\dagger3}\left|0,0\right\rangle =\sqrt{6},\label{eq:60}
\end{equation}
\begin{equation}
\left\langle 2,1\right|a_{d}^{\dagger}a_{g}^{\dagger}a_{d}^{\dagger}\left|0,0\right\rangle =\sqrt{2}\left\langle 2,1\mid2,1\right\rangle =\sqrt{2},\label{eq:61}
\end{equation}
\begin{equation}
\left\langle 1,2\right|a_{g}^{\dagger}a_{d}^{\dagger}a_{g}^{\dagger}\left|0,0\right\rangle =\sqrt{2},\label{eq:62-2}
\end{equation}
\begin{equation}
\left\langle 0,1\right|a_{d}a_{d}^{\dagger}a_{g}^{\dagger}\left|0,0\right\rangle =\left\langle 0,1\mid0,1\right\rangle =1,\label{eq:63-1}
\end{equation}
with the same method we find
\begin{equation}
\begin{array}{cc}
 & \left\langle 0,1\right|a_{d}a_{d}^{\dagger}a_{g}^{\dagger}\left|0,0\right\rangle =\left\langle 0,1\right|a_{g}^{\dagger}a_{d}a_{d}^{\dagger}\left|0,0\right\rangle \\
= & \left\langle 1,0\right|a_{d}^{\dagger}a_{g}a_{g}^{\dagger}\left|0,0\right\rangle =\left\langle 1,0\right|a_{g}a_{d}^{\dagger}a_{g}^{\dagger}\left|0,0\right\rangle \\
= & \left\langle 1,0\right|a_{g}a_{g}^{\dagger}a_{d}^{\dagger}\left|0,0\right\rangle =1.
\end{array}\label{eq:64-1}
\end{equation}

Finally, we calculate the contribution of the last term in equation
(\ref{eq:28}), i.e. $2i\hbar y^{s}$. So the first order contribution
of this term on the ground state is zero 
\begin{equation}
2i\hbar\left\langle 0,0\right|y^{s}\left|0,0\right\rangle =0.\label{eq:65-1}
\end{equation}

For the second order contribution, only $a_{d}^{\dagger}$ and $a_{g}^{\dagger}$
have non-zero contribtion 
\begin{equation}
\left\langle 1,0\right|a_{d}^{\dagger}\left|0,0\right\rangle =1,\label{eq:66-1}
\end{equation}
\begin{equation}
\left\langle 0,1\right|a_{g}^{\dagger}\left|0,0\right\rangle =1.\label{eq:67-1}
\end{equation}

One can check that the corrections due to $2i\hbar y^{s}$ on excited
states (degenerate states) to the first order is zero. Thus we find
that to the first order of perturbation, the corrections due to $H_{K}^{(\tau)}$
on the excited degenerate states vanish.

\subsection{\label{III.D}Upper bound on dynamical noncommutative parameter $\tau$}

In this subsection, we put an upper bound on the DNC parameter $\tau$
using the accuracy of the energy measurement. The second order correction
to the energy of the system is given by 
\begin{equation}
E_{n}^{(2)}=\sum_{k\neq n}\frac{\left|\left\langle \varphi_{n}\right|H_{(\tau)}\left|\varphi_{k}\right\rangle \right|^{2}}{E_{n}^{(0)}-E_{k}^{(0)}},\label{eq:68-1}
\end{equation}
but 
\begin{equation}
\left|\left\langle \varphi_{n}\right|H_{(\tau)}\left|\varphi_{k}\right\rangle \right|^{2}\propto\frac{\hbar^{2}}{\Gamma^{2}}v_{F}^{2}\tau^{2},\label{eq:69}
\end{equation}
 and 
\begin{equation}
E_{0}^{(0)}-E_{k}^{(0)}\propto\hbar\Gamma v_{F},\label{eq:70}
\end{equation}
where lower index $0$ refers to the ground state.

$E_{n}^{(2)}$ should be equal or less than the accuracy of the energy
measurement. $E_{n}^{(2)}\leq10^{-3}\text{eV}$ \cite{key-39}, thus
we have
\begin{equation}
\frac{\hbar}{\Gamma^{3}}v_{F}\tau^{2}\leq10^{-3}\text{eV}.\label{eq:71-1}
\end{equation}

Using the following numerical values of relevant quantities:  $l_{B}\approx2.5\times10^{-8}\text{ m}$
(for B=1 Tesla) ;$v_{F}\approx10^{6}m/s$ and $\hbar\approx6\times10^{-15}\text{eVs}$,
thus, the DNC parameter $\tau$ satisfies 
\begin{equation}
\sqrt{\tau}\leq\frac{1}{10^{-7}}\text{ m}^{-1}.\label{eq:72}
\end{equation}

Using the relation $1\text{ Fermi}{}^{-1}\approx200\text{ MeV}$,
one can get
\begin{equation}
\sqrt{\tau}\leq10^{-6}\text{MeV}=1\text{eV}.\label{eq:73}
\end{equation}

Clearly, this bound is not a stringment bound. But if we take the
accuracy of energy measurement $10{}^{-12}\text{eV}$ \cite{key-45,key-46}
(as in atomic physics), we obtain a much better stringment bound. 

It is worth mentioning that the upper and lower bounds on the NC parameter
$\Theta$ were obtained in \cite{key-31,key-47,key-48}.

\subsection{\label{III.E}Thermodynamic properties- Dynamics at zero temperature}

We pursue to determine the thermodynamic properties of the graphene
under a magnetic field in DNC space at zero temperature. For many
applications of interest to physicists the temperatures are \textquotedblleft small\textquotedblright ,
so the zero temperature limit is correct and valid. At $T=0$ the
ground state of a system of N fermions will have all single particle
energy levels filled up to the Fermi energy $E_{F}$ and the remainder
empty. In what follows we study the thermodynamic properties of graphene
as a relativistic Fermi system in DNC space and at $T=0$. In \cite{key-49},
the equations of state for the extreme relativistic case ($E=\left|\overrightarrow{p}\right|c$)
at $T=0$, are summarized as follows: 
\begin{equation}
n\left(p_{F}\right)=\frac{gp_{F}^{3}}{6\pi^{2}\hbar^{3}},\label{eq:81-2}
\end{equation}
\begin{equation}
u\left(p_{F}\right)=\frac{gcp_{F}^{4}}{8\pi^{2}\hbar^{3}},\label{eq:82-1}
\end{equation}
\begin{equation}
\mu=p_{F}c,\label{eq:83-1}
\end{equation}
\begin{equation}
P=\frac{1}{3}u,\label{eq:84-1}
\end{equation}
\begin{equation}
\frac{dP}{dn}=\frac{c}{3}p_{F}=\frac{1}{3}\mu,\label{eq:85-1}
\end{equation}
\begin{equation}
\gamma=\frac{4}{3},\label{eq:86-1}
\end{equation}
where $n$, $u$, $\mu$, $P$, $p_{F}$ and $g=2s+1,$ are density
of electrons, density of energy, chemical potential, pressure of the
system, Fermi momentum and degeneracy factor (that counts the number
of states available with the same momentum and position, where $s$
stands for spin), respectively. As we mentioned before, in the ground
state of the N-fermion system, the particles occupy the lowest energy
states available. Those states with energy below $E_{F}$ have unit
probability to be occupied, those with energies above $E_{F}$ remain
empty. On the other hand, it is well known fact in quantum mechanics
that, for the ground state, the second-order perturbation energy correction
term is negative. So the correction obtained in subsection \textquotedblleft \ref{III.D}\textquotedblright{}
for the ground state is negative, and therefore we have 
\begin{equation}
E_{F}^{\tau\neq0}<E_{F}^{\tau=0}.\label{eq:87-1}
\end{equation}

By considering together equations (\ref{eq:81-2}-\ref{eq:86-1})
and (\ref{eq:87-1}), we find the interesting relations between thermodynamics
properties of graphene in the DNC space and commutative one, which
are summarized as follows 

\begin{center}
{\large{}}%
\begin{tabular}{|c|}
\hline 
{\large{}$n^{\tau\neq0}<n^{\tau=0}$}\tabularnewline
\hline 
{\large{}$u^{\tau\neq0}<u^{\tau=0}$}\tabularnewline
\hline 
{\large{}$\mu^{\tau\neq0}<\mu^{\tau=0}$}\tabularnewline
\hline 
{\large{}$P^{\tau\neq0}<P^{\tau=0}$}\tabularnewline
\hline 
{\large{}$\left(\frac{dP}{dn}\right)^{\tau\neq0}>\left(\frac{dP}{dn}\right)^{\tau=0}$}\tabularnewline
\hline 
{\large{}$\gamma^{\tau\neq0}=\gamma^{\tau=0}.$}\tabularnewline
\hline 
\end{tabular}
\par\end{center}{\large \par}

It is worth mentioning that the dynamical noncommutative graphene
is more compressible than the commutative one.
\[
\left(\frac{dn}{dP}\right)^{\tau\neq0}>\left(\frac{dn}{dP}\right)^{\tau=0}.
\]

So, in summary, the behavior of graphene at zero temperature might
play an important role in understanding more the dynamical noncommutativity
and the structure of space-time.

\section{{\normalsize{}\label{IV}Conclusion}}

It is very interesting to invistegate fundamental phenomena in DNC
space. In this work, we have studied the effects of DNC space on the
graphene in the presence of an external constant magnetic field. Graphene
is described by Dirac equation so we investigate the Dirac equation
in DNCS space and derive the corrections due to dynamical noncommutativity
on the Hamiltonian and energy spectrum.  Moreover, using the accuracy
of energy measurement we set an upper bound on DNC parameter $\tau$
i.e. $\sqrt{\tau}\leq1\text{ eV}$, a bound that is not very stringent.
Of course, that indicates that there is no contradiction between DNC
effects and graphene\textquoteright s physics. Then, we investigated
some thermodynamic characteristics of the considered system at zero
temperatures. We showed that, there are significant differences between
thermodynamics properties of graphene at low temperatures in commutative
and DNC spaces. Therefore high-precision very low temperature techniques
can be used to determine whether there exists dynamical noncommutativity
of space in nature.


\begin{thebibliography}{10}
\bibitem{key-1}Geim, A., Novoselov, K. The rise of graphene. Nat.
Mater. \textbf{6}, 183 (2007). \url{https://doi.org/10.1038/nmat1849}

\bibitem{key-2}Peres, N. M. R., \& Ribeiro, R. M. Focus on graphene.
New J. Phys. \textbf{11}(9), 095002 (2009). \url{https://doi.org/10.1088/1367-2630/11/9/095002}

\bibitem{key-3}Geim, A. K. Graphene: status and prospects. science,
\textbf{324}(5934), 1530 (2009). \url{https://doi.org/10.1126/science.1158877}

\bibitem{key-4}Bonaccorso, F., Sun, Z., Hasan, T. et al. Graphene
photonics and optoelectronics. Nat. Photon. \textbf{4}, 611 (2010).
\url{https://doi.org/10.1038/nphoton.2010.186}

\bibitem{key-5}Mikhailov, S. A. \& Ziegler, K. New electromagnetic
mode in graphene. Phys. Rev. Lett. \textbf{99}, 016803 (2007). \url{https://doi.org/10.1103/PhysRevLett.99.016803}

\bibitem{key-6}Jiang, Z., et al.. Infrared Spectroscopy of Landau
Levels of Graphene. Phys. Rev. Lett.\textbf{ 98}, 197403, (2007).
\url{https://doi.org/10.1103/PhysRevLett.98.197403}

\bibitem{key-7}Ramezanali, M. R., et al. Finite-temperature screening
and the specific heat of doped graphene sheets. J. Phys. A \textbf{42},
214015 (2009). \url{https://doi.org/10.1088/1751-8113/42/21/214015}

\bibitem{key-8}Hwang, E. H. \& Das Sarma, S. Dielectric function,
screening, and plasmons in two-dimensional graphene. Phys. Rev. B
\textbf{75}, 205418 (2007). \url{https://doi.org/10.1103/PhysRevB.75.205418}

\bibitem{key-9}Neto, AH Castro, et al. The electronic properties
of graphene. Rev. Mod. Phys. \textbf{81}, 109 (2009). \url{https://doi.org/10.1103/RevModPhys.81.109} 

\bibitem{key-10}Yokoyama, T. Controllable spin transport in ferromagnetic
graphene junctions. Phys. Rev. B, \textbf{77}(7), 073413 (2008). \url{https://doi.org/10.1103/PhysRevB.77.073413} 

\bibitem{key-11}Rusanov, A. I. Thermodynamics of graphene. Surf.
Sci. Rep. \textbf{69}(4), 296 (2014).\url{https://doi.org/10.1016/j.surfrep.2014.08.003}

\bibitem{key-12}Wright, A. R., et al. Thermodynamic properties of
graphene nanoribbons under zero and quantizing magnetic fields. Microelectron.
J. \textbf{40}(4), 716 (2009). \url{https://doi.org/10.1016/j.mejo.2008.11.004}

\bibitem{key-13}Pereira, J. M., Vasilopoulos, P., \& Peeters, F.
M. Graphene-based resonant-tunneling structures. Appl. Phys. Lett,
\textbf{90}(13), 132122(2007). \url{https://doi.org/10.1063/1.2717092}


\bibitem{key-14}Rastegar Sedehi, H. R., \& Khordar, R. Investigation
of specific heat in the monolayer graphene. Iran. J. Phys. Res. \textbf{20}(2),
355 (2020). \url{https://doi.org/10.47176/ijpr.20.2.38291}

\bibitem{key-15}Hawamdeh, Mustafa M., et al. Thermodynamic properties
of graphene using the static fluctuation approximation (SFA). Can.
J. Phys. \textbf{95}(3), 211 (2017). \url{https://doi.org/10.1139/cjp-2016-0310}

\bibitem{key-16}N. Peres, Colloquium: The transport properties of
graphene: An introduction. Rev. Modern Phys. \textbf{82} (2010) 2673.
\url{https://doi.org/10.1103/RevModPhys.82.2673}

\bibitem{key-17}Fefferman, C.L., Weinstein, M.I. Wave Packets in
Honeycomb Structures and Two-Dimensional Dirac Equations. Commun.
Math. Phys. \textbf{326}, 251 (2014). \url{https://doi.org/10.1007/s00220-013-1847-2}

\bibitem{key-18}Peres N M R .The transport properties of graphene.
J. Phys.: Condens. Matter \textbf{21,} 323201 (2009). \url{https://doi.org/10.1088/0953-8984/21/32/323201}

\bibitem{key-19}Katsnelson, M. I., \& Novoselov, K. S. Graphene:
New bridge between condensed matter physics and quantum electrodynamics.
Solid State Commun. \textbf{143}(1), 3 (2007). \url{https://doi.org/10.1016/j.ssc.2007.02.043}

\bibitem{key-20}Katsnelson, M., Novoselov, K. \& Geim, A. Chiral
tunnelling and the Klein paradox in graphene. Nat. Phys. \textbf{2},
620 (2006). \url{https://doi.org/10.1038/nphys384}

\bibitem{key-21}Katsnelson, M. I. Graphene: carbon in two dimensions.
Materials Today, \textbf{10}(1), 20 (2007).. \url{https://doi.org/10.1016/S1369-7021(06)71788-6} 

\bibitem{key-22}Haouam, I. Dirac Oscillator in Dynamical Noncommutative
Space. Preprints , 2021020072 (2021). \url{https://doi.org/10.20944/preprints202102.0072.v1}

\bibitem{key-23}Andreas Fring et al. Strings from position-dependent
noncommutativity. J. Phys. A: Math. Theor. \textbf{43}, 345401 (2010).
. \url{https://doi.org/10.1088/1751-8113/43/34/345401}

\bibitem{key-24}Alavi, S.A., Rezaei, N. Dirac equation, hydrogen
atom spectrum and the Lamb shift in dynamical non-commutative spaces.
Pramana. J. Phys \textbf{88}, 77 (2017). \url{https://doi.org/10.1007/s12043-017-1381-4}

\bibitem{key-25}Gomes M and Kupriyanov V G. Position-dependent noncommutativity
in quantum mechanics. Phys. Rev. D\textbf{79}, 125011 (2009 ). \url{https://doi.org/10.1103/PhysRevD.79.125011}

\bibitem{key-26}Haouam, I. The Non-Relativistic Limit of the DKP
Equation in Non-Commutative Phase-Space. Symmetry, \textbf{11}, 223
(2019). \url{https://doi.org/10.3390/sym11020223}

\bibitem{key-27}Haouam, I. Analytical solution of (2+ 1) dimensional
Dirac equation in time-dependent noncommutative phase-space. Acta.
polytech. \textbf{60}(2), 111(2020). \url{https://doi.org/10.14311/AP.2020.60.0111}

\bibitem{key-28}Seiberg, N; Witten, E. String theory and noncommutativegeometry.
J. High Energy Phys. JHEP09, 032 (1999). \url{https://doi.org/10.1088/1126-6708/1999/09/032}

\bibitem{key-29}Gingrich, D.M. Noncommutative geometry inspired blackholes
in higher dimensions at the LHC. J. High Energ. Phys. \textbf{2010},
22 (2010). \url{https://doi.org/10.1007/JHEP05(2010)022}

\bibitem{key-30}Gracia-Bondia. J. M. Notes on quantum gravity and
noncommutative geometry: New Paths Towards Quantum Gravity.Springer,
Berlin, Heidelberg, 2010. 3-58. \url{https://doi.org/10.1007/978-3-642-11897-5_1}

\bibitem{key-31}M. Chaichian, M.M. Sheikh-Jabbari, A. Tureanu. Hydrogen
atom spectrum and the lamb shift in noncommutative QED. Phys. Rev.
Lett. \textbf{86},2716 (2001). \url{https://doi.org/10.1103/PhysRevLett.86.2716}

\bibitem{key-32}Haouam, I. On the three-dimensional Pauli equation
in noncommutative phase-space. Acta. polytech. \textbf{61}(1), 230
(2021). \url{ https://doi.org/10.14311/AP.2021.61.0230 }

\bibitem{key-33}Haouam, I. On the noncommutative geometry in quantum
mechanics. J. Phys. Stud. \textbf{24}(2), 2002 (2020).\url{ https://doi.org/10.30970/jps.24.2002}

\bibitem{key-34}Alavi, S. A. Berry's Phase in Noncommutative Spaces.
Phys. Scr.2003, \textbf{67}, 366 (2003). \url{https://doi.org/10.1238/Physica.Regular.067a00366}

\bibitem{key-35}Alavi, S. A. Lamb shift and stark effect in simultaneous
space-space and momentum-momentum noncommutative quantum mechanics
and $\theta-$deformed SU(2) algebra. Mod. Phys. Lett. A, \textbf{22}
(05), 377 (2007). \url{https://doi.org/10.1142/S0217732307018579}

\bibitem{key-36}M.R. Douglas, N.A. Nekrasov. Noncommutative field
theory. Rev. Mod. Phys. \textbf{73}, (2001) 977. \url{https://doi.org/10.1103/RevModPhys.73.977}

\bibitem{key-37}Haouam, I. On the Fisk\textendash Tait equation for
spin-3/2 fermions interacting with an external magnetic field in noncommutativespace-time.
J. Phys. Stud. \textbf{24}, 1801 (2020). \url{https://doi.org/10.30970/jps.24.1801}

\bibitem{key-38}Szabo, R. J. Quantum field theory on noncommutative
spaces. Phys. Rep. \textbf{378}(4), 207 (2003). \url{https://doi.org/10.1016/S0370-1573(03)00059-0}

\bibitem{key-39}Bastos, C., et al.. Noncommutative graphene. Int.
J. Mod. Phys. A, \textbf{28}(16), 1350064 (2013). \url{https://doi.org/10.1142/S0217751X13500644}

\bibitem{key-40}Santos, V., Maluf, R. V., \& Almeida, C. A. S. Thermodynamical
properties of graphene in noncommutative phase\textendash space. Annals
of Physics, \textbf{349}, 402 (2014). \url{https://doi.org/10.1016/j.aop.2014.07.005}

\bibitem{key-41}Boumali A. Thermodynamic properties of the graphene
in a magnetic field via the two-dimensional Dirac oscillator. Phys.
Scr. \textbf{90}, 045702 (2015). \url{https://doi.org/10.1088/0031-8949/90/4/045702}

\bibitem{key-42}Khordad, R., Rastegar Sedehi, H.R. Magnetic susceptibility
of graphene in non-commutative phase-space: Extensive and non-extensive
entropy. Eur. Phys. J. Plus \textbf{134}, 133 (2019). \url{https://doi.org/10.1140/epjp/i2019-12558-5}

\bibitem{key-43}Gomes M and Kupriyanov V G. Position-dependent noncommutativity
in quantum mechanics. Phys. Rev. D\textbf{79}, 125011 (2009). \url{https://doi.org/10.1103/PhysRevD.79.125011}

\bibitem{key-44}Fefferman, C., \& Weinstein, M. Honeycomb lattice
potentials and Dirac points. J. Am. Math. Soc. \textbf{25}(4), 1169
(2012). \url{https://doi.org/10.1090/S0894-0347-2012-00745-0}

\bibitem{key-45}Essen, L., Donaldson, R., Bangham, M. et al. Frequency
of the Hydrogen Maser. Nat.\textbf{ 229}, 110 (1971). \url{https://doi.org/10.1038/229110a0}

\bibitem{key-46}Alavi, S. A., Abbaspour, S. Dynamical noncommutative
quantum mechanics. J. Phys. A: Math. Theor. \textbf{47} 045303 (2014).
\url{https://doi.org/10.1088/1751-8113/47/4/045303}

\bibitem{key-47}Alavi, S. A. Hyperfine splitting in noncommutative
spaces. Phys. Scr. \textbf{78}, 015005 (2008). \url{https://doi.org/10.1088/0031-8949/78/01/015005}

\bibitem{key-48}Alavi, S. A., Nodeh. S . Neutrino spin oscillations
in gravitational fields in noncommutative spaces. Phys. Scr. \textbf{90},
035301 (2015). \url{https://doi.org/10.1088/0031-8949/90/3/035301}

\bibitem{key-49}Jaffe. R. L, \textquotedblleft DEGENERATE FERMION
SYSTEMS\textquotedblright , Lecture Notes 8.322, Quantum Theory II
2006, Massachusetts Institute of Technology.\end{thebibliography}
\end{document}